\documentclass[aps,twocolumn]{revtex4}

\usepackage{amsmath}
\usepackage{epsfig}

\begin{document}

\title{Remote state preparation using non-maximally entangled states}
\author{Lin Chen}\email{deteriorate@zju.edu.cn}
\author{Yi-Xin Chen}\email{yxchen@zimp.edu.cn}
\affiliation{Zhejiang Insitute of Modern Physics, Zhejiang
University, Hangzhou 310027, People's Republic of China}

\begin{abstract}
We present a scheme in which any pure qubit
$\left|\phi\right\rangle=\cos{\theta}\left|0\right\rangle+\sin{\theta}e^{i\varphi}\left|1\right\rangle$
could be remotely prepared by using minimum classical bits and the
previously shared non-maximally entangled states, on condition
that the receiver holds the knowledge of $\theta$. Several methods
are available to check the trade-off between the necessary
entanglement resource and the achievable fidelity.
\end{abstract}
\maketitle

\section{Introduction}
The elementary resources in quantum information theory are quantum
entanglement and classical communication. By means of them, one
unknown quantum state(``qubit'') could be transmitted from a
sender(``Alice'') to a receiver(``Bob''), i.e., the process of
teleportation \cite{Bennett/PRL70/1895,Zhou/PRA64/012301}, which
indicates that people have found a new way to broadcast
information and shows better prospect than the traditional
technique \cite{Chuang}. Similar to teleportation, the remote
state preparation(RSP) is assumed that Alice completely knows the
state to be prepared by Bob, who will know part of the knowledge
on this state at most ( in many situations he even knows nothing
about this state ). The essential concern for teleportation and
RSP is the trade-off between entanglement and classical
communication. It is clear that two bits of forward classical
communication and one bit of entanglement per teleported qubit are
both necessary and sufficient during the process of teleportation.
However when it turns to the RSP, whether the amount of both
quantum and classical resources could be reduced and how the
trade-off between entanglement and classical communication will
change has been checked by many authors. For instance, Pati
\cite{Pati/PRA63/014302} has shown that a qubit chosen from
equatorial or polar great circles on a Bloch sphere can be
remotely prepared with one classical bit from Alice to Bob if they
share one bit of entanglement, which implies that the  lower bound
of classical communication \cite{Holevo} is possibly reached. Many
other techniques
\cite{Bennett/PRL87/077902,Devetak/PRL87/197901,Lo/PRA62/012313,Leung/PRL90/127905,Zeng/PRA65/022316,
quant-ph/0205009} about faithful RSP have been constructed,
including both exact and asymptotical methods.

Unlike the conventional disposal, recently, Ye {\it et al.}
\cite{Ye/PRA69/022310} proposed a new scheme in which
non-maximally entangled state plays the role of quantum channel,
instead of EPR \cite{Chuang} singlet. They showed that any pure
quantum state can be faithfully prepared by using finite classical
bits and any previously shared non-maximally entangled state. An
explicit procedure is given by \cite{quant-ph/0404004}. The scheme
\cite{Ye/PRA69/022310} of many ensembles of states remotely
prepared by using minimum classical bits and previously shared
entangled state, including all the ensembles in two-dimensional
case, has been also established.

In this paper we study a RSP protocol, in which a series of
non-maximally entangled states are employed as the quantum
channel, each of which will correspond to one area where the
transmitted state lies. In section II we describe this scheme and
demonstrate that the prior fidelity expected can be achieved,
provided that enough number of entangled states is supplied and
Bob knows the content of $\theta$. In section III we provide
several techniques to reduce the entanglement resource for the
deterministic fidelity, we try to find out the lowest expense. We
compare the present work with several former techniques, in order
to represent different characteristics of RSP protocols in section
IV. Finally, we present our conclusion and some open problems.

\section{EXPLICIT SCHEME}
The protocol is characterized as follows. A pure qubit state and
its orthonormal state have the form
\begin{eqnarray}
\left|\phi\right\rangle&=&\cos{\theta}\left|0\right\rangle+\sin{\theta}e^{i\varphi}\left|1\right\rangle,\\
\left|\bar{\phi}\right\rangle&=&\sin{\theta}\left|0\right\rangle-\cos{\theta}e^{i\varphi}\left|1\right\rangle.
\end{eqnarray}
Here, two real parameters are valued in
$0\le\theta\le\frac{\pi}{2}$ and $0\le\varphi\le\ {2}\pi$, which
define the qubit $\left|\phi\right\rangle$ as a point on the Bloch
sphere \cite{Chuang}. Alice plans to transmit this state
$\left|\phi\right\rangle$ to Bob who has the knowledge of
$\theta$. Here we define that
$A_n=\frac12\arcsin[({2}q-1)^n],n=0,1,2,...$,and $q\in[\frac12,1]$
which is the expected fidelity with which Alice transmits qubit
$\left| \phi\right\rangle$ to Bob. When
$\theta\in[\frac{\pi}{4}-A_n,\frac{\pi}{4}-A_{n+1}]$, the
prior-entangled state shared by Alice and Bob is assumed like
this:
\begin{equation}
\left| \Psi _{AB}\right\rangle
=\left|0\right\rangle\left|0\right\rangle
+\tan(\frac{\pi}{4}-A_{n+1})\left|1\right\rangle\left|1\right\rangle.
\end{equation}
Notice that we don't normalize the above state for convenience and
the same reason is applicable to all following cases. As the first
step, Alice performs a unitary operation
\[
U=\left(\begin{array}{cc}
  \frac{1}{\sqrt2} & \frac{1}{\sqrt2}e^{i\varphi} \\
 \frac{1}{\sqrt2}e^{-i\varphi} & -\frac{1}{\sqrt2}
\end{array}\right),
\]
then Alice measures her particle with basis
$\{\left|0\right\rangle,\left|1\right\rangle\}$ and broadcast 1
bit to inform Bob about the result of her measurement. After
receiving the information Bob will do nothing if he gets 0 or
$\sigma_z$ if he gets 1. Therefore he could always get such state
\begin{equation}
\left|\psi\right\rangle=\left|0\right\rangle
+\tan(\frac{\pi}{4}-A_{n+1})e^{i\varphi}\left|1\right\rangle.
\end{equation}

Second, Bob performs the CNOT gate on $\left|\psi\right\rangle$
and an ancilla state
$\left|\psi_{anc}\right\rangle=\left|0\right\rangle+y\left|1\right\rangle$
\begin{eqnarray*}
U_{CNOT}\left|\psi_B\right\rangle\left|\psi_{anc}\right\rangle&=&(\left|0\right\rangle\left|0\right\rangle+{y}\left|0\right\rangle\left|1\right\rangle\\
&&+\tan(\frac{\pi}{4}-A_{n+1})e^{i\varphi}\left|1\right\rangle\left|1\right\rangle\\
&&+\tan(\frac{\pi}{4}-A_{n+1})e^{i\varphi}y\left|1\right\rangle\left|0\right\rangle)_{B,anc},
\end{eqnarray*}
where $y=a+i\sqrt{\frac{2a\cot(2A_{n+1})}{\tan(2\theta)}-a^2-1}$,
$a$ is some constant which keeps
$\frac{2a\cot(2A_{n+1})}{\tan(2\theta)}-a^2-1\ge0$
\cite{notation1}. The
\\[0.1cm]
 reduced density matrix of B then becomes
\begin{eqnarray*}
\rho_B&=&[\left|0\right\rangle+\tan(\frac{\pi}{4}-A_{n+1})e^{i\varphi}y\left|1\right\rangle]\\
&&[\left\langle0\right|+\tan(\frac{\pi}{4}-A_{n+1})e^{-i\varphi}y^\ast\left\langle1\right|]\\
&&+[y\left|0\right\rangle+\tan(\frac{\pi}{4}-A_{n+1})e^{i\varphi}\left|1\right\rangle]\\
&&[y^\ast\left\langle0\right|+\tan(\frac{\pi}{4}-A_{n+1})e^{-i\varphi}\left\langle1\right|].
\end{eqnarray*}
Using $\{\left|\phi\right\rangle,\left|\bar{\phi}\right\rangle\}$,
the basis $\{\left|0\right\rangle,\left|1\right\rangle\}$ can be
reexpressed as
\begin{eqnarray}
\left|0\right\rangle&=&\cos{\theta}\left|\phi\right\rangle+\sin{\theta}\left|\bar{\phi}\right\rangle,\\
\left|1\right\rangle&=&e^{-i\varphi}(\sin{\theta}\left|\phi\right\rangle-\cos{\theta}\left|\bar{\phi}\right\rangle).
\end{eqnarray}
So $\rho_B$ is written as
\begin{eqnarray*}
\rho_B&\equiv&C_0\left|\phi\right\rangle\left\langle\phi\right|+C_1\left|\bar{\phi}\right\rangle\left\langle\bar{\phi}\right|
+C_2\left|\phi\rangle\langle\bar{\phi}\right|+C_3\left|\bar{\phi}\rangle\langle\phi\right|,
\end{eqnarray*}
where
\begin{eqnarray*}
C_2 &=& {C_3}^\ast\\
&=&(\cos{\theta}+y\tan(\frac{\pi}{4}-A_{n+1})\sin{\theta})\\
&&(\sin{\theta}-y^\ast\tan(\frac{\pi}{4}-A_{n+1})\cos{\theta})\\
&&+(y\cos{\theta}+\tan(\frac{\pi}{4}-A_{n+1})\sin{\theta})\\
&&(y^\ast\sin{\theta}-\tan(\frac{\pi}{4}-A_{n+1})\cos{\theta})\\
&=& 0.
\end{eqnarray*}
Hence, we get
\begin{equation}
\rho_B=C_0\left|\phi\rangle\langle\phi|+C_1|\bar{\phi}\rangle\langle\bar{\phi}\right|.
\end{equation}
From the above equation, one can read off the fidelity of
$\left|\phi\right\rangle$
\begin{eqnarray}
F(\left|\phi\right\rangle\left\langle\phi\right|)=\frac{C_0}{C_0+C_1}&\equiv&\frac{1}{1+\chi},
\end{eqnarray}
where
\begin{eqnarray}
\chi&=&\frac{C_1}{C_0}=\frac{\cos{2\theta}-\sin{2A_{n+1}}}{\cos{2\theta}+\sin{2A_{n+1}}}.
\end{eqnarray}
According to $\theta\in[\frac{\pi}{4}-A_n,\frac{\pi}{4}-A_{n+1}]$,
we find
\begin{eqnarray*}
\chi_{_{\scriptstyle {MIN}}}&=& 0, \\
\chi_{_{\scriptstyle {MAX}}}&=& q^{-1}-1.
\end{eqnarray*}
It can be easily found that
$F(\left|\phi\right\rangle\left\langle\phi\right|)\in[q,1]$, which
implies that $q$ is the $minimum$ fidelity with which Bob gets
state $\left|\phi\right\rangle$.

Until now the parameter $\theta$ is confined in some smaller
region. Since $\theta\in[\frac{\pi}{4}-A_n,\frac{\pi}{4}-A_{n+1}]$
and $A_n=\frac12\arcsin[({2}q-1)^n],n=0,1,2...$, we can see that
$A_n$ will become smaller as $n$ goes up and finally
\begin{equation*}
\lim_{n\to\infty}A_n=0.
\end{equation*}
If all regions are combined ( note that $A_0=\frac{\pi}{4}$ )
\begin{eqnarray}
[\frac{\pi}{4}-A_0,\frac{\pi}{4}-A_1]\sqcup[\frac{\pi}{4}-A_1,\frac{\pi}{4}-A_2]\nonumber\\
\sqcup...[\frac{\pi}{4}-A_n,\frac{\pi}{4}-A_{n+1}]\sqcup...=[0,\frac{\pi}{4}],
\end{eqnarray}
the whole region of $[0,\frac{\pi}{4}]$ is covered. Now, if we own
sufficient non-maximally entangled states $\left| \Psi
_{AB}\right\rangle =\left|0\right\rangle\left|0\right\rangle
+\tan(\frac{\pi}{4}-A_{n+1})\left|1\right\rangle\left|1\right\rangle,n=0,1,2...$,
the protocol for $\theta\in[0,\frac{\pi}{4}]$ is completed.

On the other hand, we can deal with the region
$\theta\in[\frac{\pi}{4},\frac{\pi}{2}]$ in a similar method.
First, $\theta$ is divided into many small regions, i.e.,
$[\frac{\pi}4+A_{n+1},\frac{\pi}4+A_n],n=0,1,2...$ . Then, on each
small region, a non-maximally entangled state is provided in the
following form
\begin{equation}
\left| \Psi _{AB}\right\rangle
=\left|0\right\rangle\left|0\right\rangle +\tan(\frac{\pi}{4}+
A_{n+1})\left|1\right\rangle\left|1\right\rangle.
\end{equation}
Subsequently, the procedure is entirely the same as that
\\[0.2cm]
of region $\theta\in[0,\frac{\pi}4]$, except that $y$ will be
redefined as
$y=a+i\sqrt{-\frac{2a\cot(2A_{n+1})}{\tan(2\theta)}-a^2-1}$. Again
$a$ is some constant which keeps
$-\frac{2a\cot(2A_{n+1})}{\tan(2\theta)}-a^2-1\ge0$. After
performing
\\[0.1cm]
all steps, we get
\begin{eqnarray}
\chi&=&\frac{C_1}{C_0}=\frac{\cos{2\theta}+\sin{2A_{n+1}}}{\cos{2\theta}-\sin{2A_{n+1}}}.
\end{eqnarray}
According to
$\theta\in[\frac{\pi}4+A_{n+1},\frac{\pi}4+A_n],n=0,1,2..$, we
also get $\chi\in[0,q^{-1}-1]$, which induces
$F(\left|\phi\right\rangle\left\langle\phi\right|)\in[q,1]$.
Therefore $q$ also denotes the minimum fidelity on each region.
Since this protocol can be carried out on any region
$[\frac{\pi}{4}+A_{n+1},\frac{\pi}{4}+A_n],n=0,1,2...$, so we have
completed the scheme for region $[\frac{\pi}{4},\frac{\pi}{2}]$.
Combined with the conclusion on region $[0,\frac{\pi}{4}]$, the
explicit protocol is feasible on the whole region
$[0,\frac{\pi}{2}]$.

In the above protocol, the total classical cost we need is 1 bit.
A certain number of non-maximally entangled states is required for
distinct regions of $\theta$. The sufficient number is easily
imaginable, e.g., random astronomical number. However, it is
unclear what the necessary number is. Obviously, the smaller this
number is, the better this protocol will become. From the above
protocol we need two non-maximally entangled states
$\left|0\right\rangle\left|0\right\rangle +\tan(\frac{\pi}{4}-
A_{n+1})\left|1\right\rangle\left|1\right\rangle$ for region
$\theta\in[\frac{\pi}4-A_n,\frac{\pi}4-A_{n+1}]$ and
$\left|0\right\rangle\left|0\right\rangle +\tan(\frac{\pi}{4}+
A_{n+1})\left|1\right\rangle\left|1\right\rangle$ for region
$\theta\in[\frac{\pi}4+A_{n+1},\frac{\pi}4+A_n]$. A glancing
observation will lead to the result that these two states are
interconvertible by jointly local operation
$(\sigma_x)_A(\sigma_x)_B$ for any $n=0,1,2...$, which implies
that the family of non-maximally entangled states for the region
$[0,\frac{\pi}{4}]$ or the other family for the region
$[\frac{\pi}{4},\frac{\pi}{2}]$ will be enough for the whole
region $\theta\in[0,\frac{\pi}{2}]$. This result immediately
decreases the original number of entangled states to its half.
From this brief process we infer that the necessary entanglement
resource can be reduced. We analyze this parameter for the
explicit protocol above in appendix A, where the main result is
that we have to supply more and more shared entangled states with
increasing fidelity $q$ and approximation accuracy around the
central point $\theta=\frac{\pi}{4}$.

\section{REDUCTION OF SHARED ENTANGLEMENT}
We will show that it is possible to carry out the above protocol
with high fidelity, provided 1 cbit and some small number of
entangled states are available. According to the above argument,
it is necessary that
\begin{equation}
\lim_{n\to\infty}A_n=\lim_{n\to\infty}\frac12\arcsin[({2}q-1)^n]=0.
\end{equation}
Therefore $n$ must be very large as the minimum fidelity $q$
gradually tends to one, which says that $q$ has direct dependence
to the number of shared entangled states. The proposition from
appendix A indicates that, e.g., at least 38 non-maximally
entangled states are necessary to carry out the above protocol
under the condition $q=0.95,A_{38}\simeq0.01$. While $q$ is
required to be larger,$N$ will increase very fast. This conclusion
also shows that we need much more entanglement resource when $A_n$
is smaller, so we try to improve the qualification of the above
protocol by focusing on the small region near central point
$\theta=\frac{\pi}{4}$.

\begin{center}
{\bf IMPROVED PROTOCOL I}
\end{center}

The technique in section II supposes that Alice and Bob share the
non-maximally entangled state $\left| \Psi _{AB}\right\rangle
=\left|0\right\rangle\left|0\right\rangle
+\tan(\frac{\pi}{4}-A_n)\left|1\right\rangle\left|1\right\rangle$
when
$\theta\in[\frac{\pi}{4}-A_{n-1},\frac{\pi}{4}-A_n]\sqcup[\frac{\pi}{4}+A_n,\frac{\pi}{4}+A_{n-1}],n=0,1,2...,N$,
whose combination will be
$\theta\in[0,\frac{\pi}{4}-A_N]\sqcup[\frac{\pi}{4}+A_N,\frac{\pi}{2}]$.
Here we deal with the small central region
$\theta\in[\frac{\pi}{4}-A_N,\frac{\pi}{4}+A_N]$ by using one
maximally entangled state shared by Alice and Bob:
\begin{equation}
\left| \Psi
_{AB}\right\rangle=\frac{1}{\sqrt2}(\left|0\right\rangle\left|0\right\rangle+\left|1\right\rangle\left|1\right\rangle).
\end{equation}
First Alice performs the following unitary operation on her
particle
\[
U=\left(\begin{array}{cc}
 \cos{\theta} & \sin{\theta}e^{i\varphi} \\
 \sin{\theta}e^{-i\varphi} & -\cos{\theta}
\end{array}\right),
\]
then she measures this particle with basis
$\{\left|0\right\rangle,\left|1\right\rangle\}$ and broadcast 1
bit to Bob, who will do nothing if he gets 0 or $\sigma_z$ if he
gets 1. Thus Bob will get the following states both in 50\%\:
\begin{eqnarray}
\left|\phi\right\rangle&=&\cos{\theta}\left|0\right\rangle+\sin{\theta}e^{i\varphi}\left|1\right\rangle,\\
\left|{\phi}^\prime\right\rangle&=&\sin{\theta}\left|0\right\rangle+\cos{\theta}e^{i\varphi}\left|1\right\rangle.
\end{eqnarray}
So if Bob receives 0 the protocol comes true, while he gets state
$\left|{\phi}^\prime\right\rangle$, we observe that
$\theta\in[\frac{\pi}{4}-A_N,\frac{\pi}{4}+A_N]$ and if $A_N$ is
very small, the two qubits $|\phi\rangle$ and
$|\phi^\prime\rangle$ will be very close to each other. Hence, the
fidelity is
\begin{eqnarray}
F(|\phi\rangle,|\phi^\prime\rangle)\equiv|\langle\phi^\prime|\phi\rangle|=\sin{2\theta},
\end{eqnarray}
so we obtain
\begin{eqnarray}
F\ge F_{min}=\sin(\frac{\pi}{2}\pm2A_N)=\sqrt{1-(2q-1)^{2N}}.
\end{eqnarray}
The connection between $N$ and $q$ is described in figure 1 and
figure 2, while several key points $N+1$ in the following table
for there is one added maximally-entangled state. This protocol is
a kind of $approximate$ substitution, which indeed requires that
qubit $|{\phi}^\prime\rangle$ replaces $|\phi\rangle$ with a high
fidelity. We also give another scheme based on the above
discussion in appendix B, which is a kind of probabilistic exact
protocol \cite{quant-ph/0307100}. Combined with the technique in
section II on region
$[0,\frac{\pi}{4}-A_N]\sqcup[\frac{\pi}{4}+A_N,\frac{\pi}{2}]$,
the whole protocol is now completed.
\\[0.1cm]

\begin{figure}[ht]
\epsfxsize=7.8cm \epsfbox{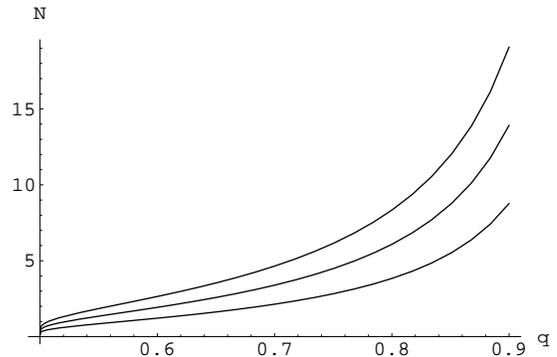} \caption{trade-off
between $N$ and $q \in [0.5,0.9]$. The three curves from downside
to upside represent $F_{min}=0.99,0.999,0.9999$, whose precision
is gradually improved. $N$ increases very slowly.}
\end{figure}

\begin{figure}[ht]
\epsfxsize=7.8cm \epsfbox{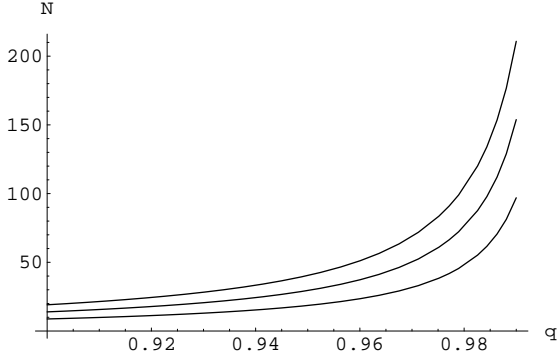} \caption{trade-off
between $N$ and $q \in [0.9,0.99]$. The three curves from downside
to upside represent $F_{min}=0.99,0.999,0.9999$. We find $N$ here
increases less than that of Appendix A.}
\end{figure}

\begin{center}
\begin{tabular}{|l|l|l|l|l|} \hline
{$F_{min}$}&{\bf$q=0.90$}&{\bf$q=0.95$}&{\bf$q=0.98$}&{\bf$q=0.99$}\\\hline
0.99&9.78&19.59&48.98&97.94\\\hline
0.999&14.93&30.49&77.12&154.82\\\hline
0.9999&20.08&41.42&105.32&211.80\\\hline
\end{tabular}
\end{center}

\begin{center}
{\bf IMPROVED PROTOCOL II}
\end{center}
We take an ulterior step to decrease the entanglement resource.
The present idea is based on the technique in appendix A, which
leaves out the small symmetry region
$[\frac{\pi}{4}-A_N,\frac{\pi}{4}+A_N]$. We define
$C_n=\tan(\frac{\pi}{4}-A_n)$, thus $N$ necessary entangled states
are $\left| \Psi_{AB}\right\rangle
=\left|0\right\rangle\left|0\right\rangle
+C_{l_k}\left|1\right\rangle\left|1\right\rangle,
1={l_0}\le{l_1}\le...\le{l_{k-1}}\le{l_k}\le...\le{l_{M-1}}\le{l_M}=N$.

Firstly, we divide these $N$ states into $M$ sections
\begin{equation}
\left|
{\Psi_k}\right\rangle=\left|0\right\rangle\left|0\right\rangle
+C_{f_k}\left|1\right\rangle\left|1\right\rangle,f_k\in[l_{k-1},l_k],k\in[1,M].
\end{equation}
Evidently, each section contains several entangled states. Our aim
is to replace $\left|{\Psi_k}\right\rangle$ by
\begin{equation}
\left|
{\Phi_k}\right\rangle=\left|0\right\rangle\left|0\right\rangle
+B_k\left|1\right\rangle\left|1\right\rangle,k\in[1,M].
\end{equation}
$B_k$ is a positive constant. We introduce a POVM measurement
\begin{eqnarray}
M_{k0}&=&\left(\begin{array}{cc}
\sqrt{\frac{{B_k}^2+1}{{C_{f_k}}^2+1}}\sqrt{P_k} & 0\\
 0
 &\sqrt{\frac{{B_k}^2+1}{{C_{f_k}}^2+1}}\sqrt{P_k}\frac{C_{f_k}}{B_k}\end{array}\right),\nonumber\\
M_{k1}&=&\sqrt{I-M_{k0}^2} ,
\end{eqnarray}
where $P_k\in[\frac{1}{{B_k}^2+1},1]$,
$C_{f_k}\in[\sqrt{{P_k}{B_k}^2+P_k-1},B_k]$. After performing
measurement $M_{k0}$, therefore $\left| {\Phi_k}\right\rangle$ can
be transformed into $\left| {\Psi_k}\right\rangle$. The
probability Alice carries out $M_{k0}$ is
\begin{eqnarray}
P(M_{k0})=\left\langle{\Phi_k}\right|M_{k0}^\dagger
M_{k0}\left|{\Phi_k}\right\rangle=P_k.
\end{eqnarray}
The above argument implies that Alice can decrease the necessary
number of entangled states with probability $P_k$, by substituting
one state $\left|{\Phi_k}\right\rangle$ for each section $\left|
{\Psi_k}\right\rangle$.

Here we give a concrete procedure to show how entanglement
resource is reduced. According to the result in appendix A, at
least $N=194$ entangled states are required so that the small
region $[\frac{\pi}{4}-A_N,\frac{\pi}{4}+A_N]$ can be left out
with $q=0.99$. For convenience we suppose $P_k=0.99$ for
$k\in[1,M]$. The necessary condition to be satisfied is
$C_{f_k}\in[\sqrt{{P_k}{B_k}^2+P_k-1},B_k]$, thus all $f_k$
belonging to this region will lead to the fact that $\left|
{\Psi_k}\right\rangle=\left|0\right\rangle\left|0\right\rangle
+C_{f_k}\left|1\right\rangle\left|1\right\rangle$ is replaced by
$\left|
{\Phi_k}\right\rangle=\left|0\right\rangle\left|0\right\rangle
+B_k\left|1\right\rangle\left|1\right\rangle$. First we set
$B_M=\tan(\frac{\pi}{4}-A_{194})$, i.e., $l_M=194$. In order to
get $l_{M-1}$ we calculate
\begin{equation}
\sqrt{0.99\times{\tan(\frac{\pi}{4}-A_{194})}^2+0.99-1}
\leq{C_{f_k}}\leq{\tan(\frac{\pi}{4}-A_{194})},
\end{equation}
thus
\begin{equation}
173.312\leq{f_k}\leq194.
\end{equation}
So we obtain $l_{M-1}=174$. The next aim is to find out $l_{M-2}$.
It is noticed that the end $l_{M-1}=174$ has been included in the
first region $[l_{M-1},l_M]$, and in fact this state needs not to
be included in the second region. Without loss of generality we
still adopt the mark $f_k\in[l_{k-1},l_k],k\leq{M-1}$, where the
point $f_k=l_k$ indeed belongs to the former region
$[l_k,l_{k+1}]$. So we set $B_{M-1}=\tan(\frac{\pi}{4}-A_{173})$
to get
\begin{equation}
\sqrt{0.99\times{\tan(\frac{\pi}{4}-A_{173})}^2+0.99-1}
\leq{C_{f_k}}\leq{\tan(\frac{\pi}{4}-A_{173})},
\end{equation}
which leads to $l_{M-2}=159$. The technique for rest region is
analogous to the above procedure. At last the total number of
entangled states
$\left|{\Phi_k}\right\rangle=\left|0\right\rangle\left|0\right\rangle
+\tan(\frac{\pi}{4}-A_k)\left|1\right\rangle\left|1\right\rangle$
is $M=50:k=194,173,158,146,136,128,121,115,109,104,99,94,90,86$,
$82,78,75,72,69,66,63,60,57,2t,1\leq{t}\leq27$. Notice that
$P_k\geq\frac{1}{{B_k}^2+1}$ for all $k$ above. If lower
probability is allowed, e.g., $P_k=0.98$ for $k\in[1,M]$, similar
technique says that only $M=29$ entangled states are required.

Now we summarize the whole protocol. First Alice and Bob share $M$
quantum channels $\left|
{\Phi_k}\right\rangle=\left|0\right\rangle\left|0\right\rangle
+B_k\left|1\right\rangle\left|1\right\rangle,k\in[1,M]$. By local
POVM measurement $\{M_{k0},M_{k1}\}$, Alice can transform each
$\left| {\Phi_k}\right\rangle$ into corresponding string $\left|
{\Psi_k}\right\rangle=\left|0\right\rangle\left|0\right\rangle
+C_{f_k}\left|1\right\rangle\left|1\right\rangle,
f_k\in[l_{k-1},l_k], k\in[1,M]$, with probability $P_k$. Here, we
define $l_0=1$ and $l_M=N$. $N$ is determined by $q$ and the
approximation accuracy of region
$[\frac{\pi}{4}-A_N,\frac{\pi}{4}+A_N]$. Next step follows the
technique in section II, since we have got $\left| \Psi
_{AB}\right\rangle =\left|0\right\rangle\left|0\right\rangle
+\tan(\frac{\pi}{4}-A_k)\left|1\right\rangle\left|1\right\rangle,k\in[1,N]$.
 The assumption Bob knows $\theta$ assists Bob by distinguishing
which channel is in use, i.e., state $\left|
{\Phi_k}\right\rangle$ corresponds to
$\theta\in[\frac{\pi}{4}-A_{l_{k-1}-1},\frac{\pi}{4}-A_{{l_k}-1}]\sqcup[\frac{\pi}{4}+A_{{l_k}-1},\frac{\pi}{4}+A_{l_{k-1}-1}]$
when $1\leq{k}\leq{M-1}$, and
$\theta\in[\frac{\pi}{4}-A_{l_{M-1}-1},\frac{\pi}{4}-A_{{l_M}}]\sqcup[\frac{\pi}{4}+A_{{l_M}},\frac{\pi}{4}+A_{l_{M-1}-1}]$
when $k=M$. Finally Bob will get the expected state
$\left|\phi\right\rangle$ with a minimum fidelity $q\times{P_k}$.

Hitherto we construct protocol II based on the demonstration in
appendix A. However, it is completely feasible to adopt the
technique in protocol I and appendix B for the disposal of region
$[\frac{\pi}{4}-A_N,\frac{\pi}{4}+A_N]$. Combined with appendix B,
a probabilistic exact protocol with higher efficiency is
practicable. If more entanglement resource is available, we can
improve the success probability farther.

\section{MORE ARGUMENT ABOUT RSP}

All techniques above describe a sort of RSP protocol with a
decided fidelity, which requires the expense of certain number of
entangled states and one bit of classical communication. An
apparent deficiency in this protocol is that the receiver needs to
know the content of $\theta$, and the entanglement resource
required may be large. The technique based on the dark states
\cite{quant-ph/0201138,quant-ph/0304006} provided an explicit
scheme in which Bob knows $\theta$ or $\varphi$, under this
condition Alice could transmit any qubit
$\left|\phi\right\rangle=\cos{\theta}\left|0\right\rangle+\sin{\theta}e^{i\varphi}\left|1\right\rangle$
to Bob with the expense of one maximally-entangled state and one
bit of classical communication. However the transformation it
requires in the case Bob knows $\theta$ is not unitary but
Hermitian, therefore it is not possible to carry out such protocol
on a quantum computer \cite{Chuang}. The probabilistic exact
protocol in \cite{quant-ph/0302170} could transmit any polar state
with relatively lower fidelity from one sender to different
receivers, where the expense is one deliberate entangled state and
one cbit.

Another idea from \cite{Ye/PRA69/022310} has given a faithful
scheme in which many ensembles of states can be remotely prepared
by using minimum classical bits and previously shared entangled
state, especially they have found all the ensembles in
two-dimensional case. It seems this is a better protocol for it
needs only one shared entangled state and one cbit, furthermore
this is a faithful scheme. Here we do some simple analysis on this
scheme in two-dimensional case. As described in
\cite{Ye/PRA69/022310}, the ensemble that can be remotely prepared
must be in the form
\begin{equation}
\left\{ v\left| \Phi \right\rangle =v\left( \alpha _0\left|
0\right\rangle +\alpha _1e^{i\omega }\left| 1\right\rangle \right)
,\alpha _0,\alpha _1>0,\alpha _0^2+\alpha _1^2=1,\forall \omega
\right\}
\end{equation}
by a previously shared entangled state
\begin{equation}
\left| \Psi _{AB}\right\rangle
=\alpha_0\left|0\right\rangle\left|0\right\rangle
+\alpha_1\left|1\right\rangle\left|1\right\rangle.
\end{equation}
Here, $\alpha _0$,$\alpha _1$ and $\omega$ are known to Alice and
$v$ is a unitary operator in two-dimensional Hilbert space.
Therefore we suppose
\[
v=\left(\begin{array}{cc}
  \cos{\gamma} & -e^{i\delta}\sin{\gamma} \\
 e^{i\beta}\sin{\gamma} & e^{i(\beta+\delta)}\cos{\gamma}
\end{array}\right),
\]
where $\beta$,$\gamma$,$\delta$ are real parameters. This
operation is done by Bob so the parameters should be independent
of $|\phi\rangle$. Since all the ensembles have been found, we can
infer
\begin{equation}
v\left|\Phi\right\rangle=Ae^{-i\alpha}\left|\phi\right\rangle=Ae^{-i\alpha}(\cos{\theta}\left|0\right\rangle+\sin{\theta}e^{i\varphi}\left|1\right\rangle),
\end{equation}
where $A$ and $\alpha$ are any real number. Obviously $A=\pm1$,
and if $A=-1$ we can set $\gamma\to\gamma+\pi$. Thus
\begin{equation}
v\left|\Phi\right\rangle=e^{-i\alpha}(\cos{\theta}\left|0\right\rangle+\sin{\theta}e^{i\varphi}\left|1\right\rangle).
\end{equation}
Some simple algebra will lead to
\begin{eqnarray*}
\alpha_0&=&\sqrt{\cos^2{\gamma}-\cos{2\gamma}\sin^2{\theta}+\frac12\sin{2\gamma}\sin{2\theta}\cos(\varphi-\beta)}\\
\alpha_1&=&\sqrt{\cos^2{\gamma}-\cos{2\gamma}\cos^2{\theta}-\frac12\sin{2\gamma}\sin{2\theta}\cos(\varphi-\beta)}\\
\end{eqnarray*}
where we have employed the assumption that $\alpha _0,\alpha
_1>0,\alpha _0^2+\alpha _1^2=1$. A simple observation shows that
both $\alpha _0$ and $\alpha _1$ must be related to $\theta$ or
$\varphi$ under the assumption that $\beta$ and $\gamma$ are
constant. Therefore the shared entangled state $\left| \Psi
_{AB}\right\rangle
=\alpha_0\left|0\right\rangle\left|0\right\rangle
+\alpha_1\left|1\right\rangle\left|1\right\rangle$ $can't~be~
constant$, i.e., it is a variable which transforms with the change
of $\theta$. That is to say, $infinite$ amount of entangled states
are required to perform this protocol, for there are infinite
number of $\theta$ during the region $[0,\frac{\pi}{2}]$. However,
is it possible that $\alpha _0$ and $\alpha _1$ become constant,
provided $\beta$ and $\gamma$ are related to $\theta$ or $\varphi$
? We simply rewrite the expression of $\alpha _0$ and $\alpha _1$
by employing the two ends $\theta=0,\frac{\pi}{2}$ to get
\begin{eqnarray}
\alpha _0=\alpha _1=|\cos\gamma|=|\sin\gamma|=\frac{1}{\sqrt2},\nonumber\\
\beta=\varphi\pm\frac{\pi}{2},\omega=\pi-\delta-2\theta.
\end{eqnarray}
Therefore Bob has to know $\varphi$, in addition the shared state
$\left| \Psi _{AB}\right\rangle$ has become an EPR singlet, which
breaches the origin thought in \cite{Ye/PRA69/022310}. It is a
trivial scheme similar to that in \cite{quant-ph/0304006}. To say
the least, $\alpha _0$ and $\alpha _1$ will still be connected
with $\theta$ or $\varphi$, while $\beta$ and $\gamma$ are merely
related to $\theta$. It is because that the term
$\frac12\sin{2\gamma}\sin{2\theta}\cos(\varphi-\beta)$ will not
disappear until $\sin{2\gamma}=0$, under which the shared
entangled state remains $\left| \Psi _{AB}\right\rangle
=\cos\theta\left|0\right\rangle\left|0\right\rangle
+\sin\theta\left|1\right\rangle\left|1\right\rangle$ or $\left|
\Psi _{AB}\right\rangle
=\sin\theta\left|0\right\rangle\left|0\right\rangle
+\cos\theta\left|1\right\rangle\left|1\right\rangle$. From these
reasons we infer that the protocol in \cite{Ye/PRA69/022310}
always requires one entangled state for one corresponding
$\theta$, otherwise it will be a trivial scheme. Therefore
infinite entanglement resource is required for all transmitted
qubits. Furthermore, the condition Bob holds the knowledge of
$\theta$ will not help decrease the necessary shared entanglement.
In fact, the protocol in this paper is an effective method in
economizing entanglement, by the help that Bob knows $\theta$.

One latest technique in \cite{quant-ph/0404004} is an exactly
faithful RSP protocol, which requires finite cbits and one
entangled state in $d$-dimensional Hilbert space. This technique
requires relatively more classical communication than other
techniques, however it needs only one arbitrary entangled state to
transmit any one qubit, which is established on the transformation
of original entangled state. The idea in improved protocol II is
based on this scheme. If two forward cbits are allowed, the scheme
in this paper may become another form like this. The quantum
channel shared by Alice and Bob is one Greenberger-Horne-Zeilinger
(GHZ) state
\begin{equation}
\left| \Psi\right\rangle
=\frac{1}{\sqrt2}(\left|0\right\rangle\left|0\right\rangle\left|0\right\rangle
+\left|1\right\rangle\left|1\right\rangle\left|1\right\rangle).
\end{equation}
Here, Alice and Bob have two and one particles respectively. First
Alice performs a local unitary operation on one of her particles
\[
U=\left(\begin{array}{cc}
  \cos\theta & -\sin\theta \\
 \sin\theta & \cos\theta
\end{array}\right).
\]

Then Alice measures this particle with basis
$\{\left|0\right\rangle,\left|1\right\rangle\}$ and broadcast 1
bit to inform Bob about the result $k$ of her measurement. The
corresponding operations done by Alice and Bob are respectively
\begin{eqnarray}
U_{A0}=\left(\begin{array}{cc}
  1 & 0 \\
 0 & -e^{i\varphi}
\end{array}\right),
U_{B0}=\left(\begin{array}{cc}
  1 & 0 \\
 0 & 1
\end{array}\right), k=0
\end{eqnarray}
and
\begin{eqnarray}
U_{A1}=\left(\begin{array}{cc}
  0 & 1 \\
 e^{i\varphi} & 0
\end{array}\right),
U_{B1}=\left(\begin{array}{cc}
  0 & 1 \\
 1 & 0
\end{array}\right), k=1.
\end{eqnarray}
Thus they will always share
\begin{equation}
\left|\psi\right\rangle=\cos\theta\left|0\right\rangle\left|0\right\rangle
+\sin\theta{e^{i\varphi}}\left|1\right\rangle\left|1\right\rangle.
\end{equation}
Second, Alice performs a Hadamard gate and broadcast 1 bit to Bob,
who does nothing if he gets 0 or $\sigma_z$ if he gets 1. Now Bob
gets the expected state
$\left|\phi\right\rangle=\cos{\theta}\left|0\right\rangle+\sin{\theta}e^{i\varphi}\left|1\right\rangle$.
This protocol is also a typical application which employs the idea
of entanglement transformation.

In general, all kinds of RSP protocols try to find out the best
trade-off between the classical communication and entanglement
resource under different preconditions. As we have compared
various RSP protocols, the technique advanced in this paper widens
the traditional restriction, i.e., the receiver Bob owns the
content of parameter $\theta$ of a general qubit
$\left|\phi\right\rangle=\cos{\theta}\left|0\right\rangle+\sin{\theta}e^{i\varphi}\left|1\right\rangle$.
This is indeed a generalization of Pati's RSP protocol
\cite{Pati/PRA63/014302} by loosing the range of $\theta$.
Moreover, we also provide several techniques to decrease the
necessary entanglement resource effectively. Since the lower bound
of the entanglement consumption remains to be found, we may expect
there should be better idea to reduce the necessary resource,
following the technique in this paper. On the other hand, we also
employ the non-maximally entangled state as the quantum channel
like that in \cite{Ye/PRA69/022310,quant-ph/0404004}. The
non-maximally state is more flexible and it expands our choice in
the aspect of quantum connection. However, the generally faithful
RSP protocol is still hard to establish if the condition on
quantum and classical resources is restricted. Although the
asymptotic techniques, e.g.,
\cite{Bennett/PRL87/077902,Devetak/PRL87/197901,quant-ph/0307100}
have successfully transmitted arbitrary state at a cost of 1 cbit
and 1 ebit per qubit sent, their preconditions required are plenty
of shared entanglement resource and classical communication.

\section{Conclusions}

We have given an explicit protocol for performing the RSP
protocol, using minimum classical bits and certain number of
non-maximally entangled states as the quantum channel. The
trade-off between the necessary entanglement resource and the
achievable fidelity is discussed in detail by several different
techniques. The evaluation of this protocol should be focused on
how far we can reduce entanglement resource, however the optimal
choice is hard to make out despite the above discussion. One
useful finding in this paper is that the condition Bob knows
$\theta$ will be helpful to transmit the qubit with lower
necessary entanglement. We may consider, that there is some latent
connection between the necessary resource and how far the receiver
owns the knowledge of the qubit sent, during an exact RSP process.
The idea of entanglement transformation is a good method which may
lead to a better effect later.

\begin{center}
{\bf APPENDIX A:RESOURCE FOR THE EXPLICIT PROTOCOL}
\end{center}

It is defined that $A_n=\frac12\arcsin[({2}q-1)^n],n=0,1,2,...,N$
and $q\in[\frac12,1]$. The task we face is how to make $A_N$ as
small as possible, so that the region
$[\frac{\pi}{4}-A_N,\frac{\pi}{4}]$ can be ignored, in other words
this region has been $concentrated$ on the point
$\theta=\frac{\pi}{4}-A_N$. In this case qubit
$\left|\phi\right\rangle=\cos{\theta}\left|0\right\rangle+\sin{\theta}e^{i\varphi}\left|1\right\rangle$
will be very close to the polar state
$\cos{\frac{\pi}{4}}\left|0\right\rangle+\sin{\frac{\pi}{4}}e^{i\varphi}\left|1\right\rangle$
during the region $[\frac{\pi}{4}-A_N,\frac{\pi}{4}]$ . The second
half is similarly treated with, i.e., as $A_N$ is very small,
region $[\frac{\pi}{4},\frac{\pi}{4}+A_N]$ will be concentrated on
the point $\theta=\frac{\pi}{4}+A_N$. The estimation procedure is
below.

Firstly, to the first half we suppose
\begin{equation*}
\frac{\sin(\frac{\pi}{4}-A_N)}{\sin{\frac{\pi}{4}}}=1-10^{-m},m=2,3,4,5,6...,
\end{equation*}
i.e., $A_N=\frac{\pi}{4}-\arcsin[\frac{1-10^{-m}}{\sqrt2}]$, which
also keeps
\begin{equation*}
\frac{\cos(\frac{\pi}{4}-A_N)}{\cos{\frac{\pi}{4}}}<1+10^{-m},m=2,3,4,5,6...,
\end{equation*}
consequently the qubit
$\cos(\frac{\pi}{4}-A_N)\left|0\right\rangle+\sin(\frac{\pi}{4}-A_N)e^{i\varphi}\left|1\right\rangle$
could approximately substitute the qubit
$\cos{\theta}\left|0\right\rangle+\sin{\theta}e^{i\varphi}\left|1\right\rangle$
with high fidelity, if $m$ is large enough. Secondly, the above
procedure also applies to the second half. Since we already have
\begin{eqnarray*}
\frac{\sin(\frac{\pi}{4}+A_N)}{\cos{\frac{\pi}{4}}}<1+10^{-m},m=2,3,4,5,6...,\\
\frac{\cos(\frac{\pi}{4}+A_N)}{\sin{\frac{\pi}{4}}}=1-10^{-m},m=2,3,4,5,6...,
\end{eqnarray*}
it implies that qubit
$\cos(\frac{\pi}{4}+A_N)\left|0\right\rangle+\sin(\frac{\pi}{4}+A_N)e^{i\varphi}\left|1\right\rangle$
is also close to the polar qubit
$\cos{\frac{\pi}{4}}\left|0\right\rangle+\sin{\frac{\pi}{4}}e^{i\varphi}\left|1\right\rangle$.
According to the above argument we describe the relationship
between $N$ and $q$ in Figure 3 where
\begin{equation*}
N=\frac{\log(2*10^{-m}-10^{-2m})}{\log(2q-1)}.
\end{equation*}
Here $\log(x)$ denotes logarithms to base 2. Notice that $q$ is
the minimum fidelity for Bob to get the state $|\phi\rangle$
during the region
$[0,\frac{\pi}{4}-A_N]\sqcup[\frac{\pi}{4}+A_N,\frac{\pi}{2}]$.
Several necessary numbers $N$ of non-maximally entangled states
are explicitly provided in the following table.
\\[0.1cm]
\begin{figure}[ht]
\epsfxsize=7.8cm \epsfbox{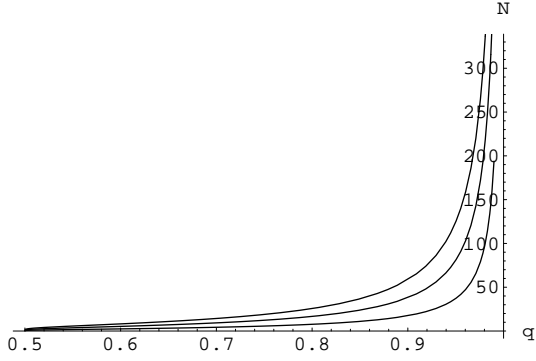} \caption{trade-off
between $N$ and $q \in [0.5,0.99]$. The three curves from downside
to upside represent $m=2,4,6$, whose precision is gradually
improved. When $q$ is low the necessary number $N$ increases very
slowly while it suddenly increases a lot as $q$ tends to one.}
\end{figure}

\begin{center}
\begin{tabular}{|l|l|l|l|l|l|}\hline
{m}&{$A_N$}&{\bf$q=0.90$}&{\bf$q=0.95$}&{\bf$q=0.98$}&{\bf$q=0.99$}\\\hline
2&$9.95066\times10^{-3}$&17.55&37.18&95.95&193.89\\\hline
4&$9.99950\times10^{-5}$&38.17&80.84&208.64&421.59\\\hline
6&$9.99999\times10^{-7}$&58.81&124.55&321.45&649.54\\\hline
\end{tabular}
\end{center}

\begin{center}
{\bf APPENDIX B: ANOTHER IMPROVED PROTOCOL}
\end{center}
We deal with the region
$\theta\in[\frac{\pi}{4}-A_N,\frac{\pi}{4}+A_N]$ in another way
which is based on correct protocol one, until Bob gets states
$|\phi\rangle$ and $|{\phi}^\prime\rangle$ both in 50\%\ . If he
gets $|{\phi}^\prime\rangle$ then Bob carries out one $POVM$
measurement
\[
M_0=\left(\begin{array}{cc} 1 & 0\\
0 & \tan^2\theta\end{array}\right),
M_1=\left(\begin{array}{cc} 0 & 0\\
0 & \sqrt{1-\tan^4\theta}\end{array}\right),
\]
if $\theta\in[\frac{\pi}{4}-A_N,\frac{\pi}{4}]$ or
\[
{M_0}^\prime=\left(\begin{array}{cc} \cot^2\theta & 0\\
0 & 1\end{array}\right),
{M_1}^\prime=\left(\begin{array}{cc} \sqrt{1-\cot^4\theta} & 0\\
0 & 0\end{array}\right),
\]
if $\theta\in[\frac{\pi}{4},\frac{\pi}{4}+A_N]$. The probabilities
with which Bob performs $M_0$ and ${M_0}^\prime$ are
\begin{eqnarray*}
P(M_0)=\left\langle{\phi}^\prime\right|M_0^\dagger
M_0\left|{\phi}^\prime\right\rangle=\tan^2\theta,
\end{eqnarray*}
\begin{eqnarray*}
P({M_0}^\prime)=\left\langle{\phi}^\prime\right|{{M_0}^\prime}^\dagger
{M_0}^\prime\left|{\phi}^\prime\right\rangle=\cot^2\theta.
\end{eqnarray*}

Evidently both $M_0$ and ${M_0}^\prime$ will transform state
 $|{\phi}^\prime\rangle$ into $|\phi\rangle$. As $A_N$ is a small amount, we may infer
Bob will do $M_0$ or ${M_0}^\prime$ with a high probability. On
the other hand Bob always gets the expected state $|\phi\rangle$
with probability 50\%\ . Therefore the fidelity is
\begin{eqnarray*}
F(\theta\in[\frac{\pi}{4}-A_N,\frac{\pi}{4}])=\frac{1+\tan^2\theta}{2}=\frac{\sec^2\theta}{2},\\
F(\theta\in[\frac{\pi}{4},\frac{\pi}{4}+A_N])=\frac{1+\cot^2\theta}{2}=\frac{\csc^2\theta}{2},
\end{eqnarray*}
which implies
\begin{equation*}
F(\theta\in[\frac{\pi}{4}-A_N,\frac{\pi}{4}+A_N])_{min}=\frac{1}{(2q-1)^N+1},
\end{equation*}
or another form
\begin{equation*}
N=\frac{\log(\frac{1}{F_{min}}-1)}{\log(2q-1)}.
\end{equation*}
This function is described in Figure 4. From it we find that $N$
increases slowly in a majority of region while $N$ will become
very large as $q$ tends to 1, which is similar to the situation in
improved protocol I. Here we also need only $N+1$ entangled states
and several key point $N+1$ is made out in the table below. The
fidelity in Figure 4 is lower than that of improved protocol I.
However it is noticeable that this protocol is a probabilistic
exact protocol, i.e., combined with the argument in section II Bob
could get the explicit qubit $|\phi\rangle$ during the whole
region $\theta\in[0,\frac{\pi}{2}]$ with a high fidelity.
Therefore such protocol is more assuring than improved protocol I
and it can also be combined with the explicit scheme in section II
to get a better qualification.

\begin{figure}[ht]
\epsfxsize=7.8cm \epsfbox{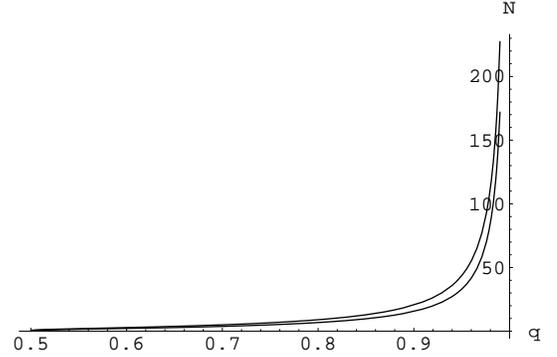} \caption{trade-off
between $N$ and $q \in [0.5,0.99]$. The two curves from downside
to upside represent $F_{min}=0.97$ and $F_{min}=0.99$.}
\end{figure}

\begin{center}
\begin{tabular}{|l|l|l|l|l|}\hline
{$F_{min}$}&{\bf$q=0.90$}&{\bf$q=0.95$}&{\bf$q=0.98$}&{\bf$q=0.99$}\\\hline
0.97&16.58&33.99&86.15&173.06\\\hline
0.99&21.59&44.61&113.57&228.45\\\hline
\end{tabular}
\end{center}

\begin{center}
{\bf Acknowledgments}
\end{center}

We thank D. Yang for his helpful comments. The work was partly
supported by the NNSF of China (Grant No.90203003), NSF of
Zhejiang Province (Grant No.602018), and by the Foundation of
Education Ministry of China (Grant No.010335025).


\begin{references}

\bibitem{Bennett/PRL70/1895}  C. H. Bennett, G. Brassard, C. Cr\'{e}peau, R.
Jozsa, A. Peres, and W. K. Wootters, Phys. Rev. Lett. {\bf 70},
1895 (1993).

\bibitem{Zhou/PRA64/012301}  J. D.  Zhou, G. Hou, and Y. D. Zhang,
Phys. Rev. A {\bf 64}, 012301 (2001).

\bibitem{Chuang}  M. A. Nielsen and I. L. Chuang, {\it Quantum Computation
and Quantum Information} (Cambridge University Press, Cambridge,
England, 2000).

\bibitem{Pati/PRA63/014302}  A. K. Pati, Phys. Rev. A {\bf 63}, 014302
(2000).

\bibitem{Holevo}  A. S. Holevo, Probl. Inf. Transm. (USSR), 9:117, 1973.

\bibitem{Bennett/PRL87/077902}  C. H. Bennett, D. P. DiVincenzo,
P. W. Shor, J.A. Smolin, B. M. Terhal, and W. K. Wootters, Phys.
Rev. Lett. {\bf 87}, 077902 (2001).

\bibitem{Devetak/PRL87/197901}  I. Devetak and T. Berger, Phys. Rev. Lett.
{\bf 87}, 197901 (2001).

\bibitem{Lo/PRA62/012313}  H. K. Lo, Phys. Rev. A {\bf 62}, 012313 (2000).

\bibitem{Leung/PRL90/127905}  D. W. Leung and P. W. Shor, Phys. Rev. Lett.
{\bf 90}, 127905 (2003).

\bibitem{Zeng/PRA65/022316}  B. Zeng and P. Zhang, Phys. Rev. A {\bf 65},
022316 (2002).

\bibitem{quant-ph/0205009}  A. Hayashi, T. Hashimoto and M. Horibe, Phys.
Rev. A {\bf 67}, 052302 (2003).

\bibitem{Ye/PRA69/022310}  M. Y. Ye, Y. S. Zhang, and G. C. Guo,
Phys. Rev. A {\bf 69}, 022310 (2004).

\bibitem{quant-ph/0404004} D. W. Berry, quant-ph/0404004.

\bibitem{notation1}  The
denominator $\tan{2\theta}$ will not affect the two ends
$\theta=0,\frac{\pi}{2}$ because we can easily set, e.g.,
$a=\frac{\tan{2\theta}}{\cot{2A_{n+1}}}$. Then
$y=\frac{\tan{2\theta}}{\cot{2A_{n+1}}}+i\sqrt{2-(\frac{\tan{2\theta}}{\cot{2A_{n+1}}})^2-1}$,
in which $\theta=0,\frac{\pi}{2}$ can be consistently defined.

\bibitem{quant-ph/0307100}  C. H. Bennett, P. Hayden, D. Leung, P. W. Shor,
A. Winter, quant-ph/0307100.

\bibitem{quant-ph/0201138} P.Kok, K.Nemoto and W.J.Munro,
quant-ph/0201138.

\bibitem{quant-ph/0304006} P.Agrawal,P.parashar and A.K.Pati,
quant-ph/0304006.

\bibitem{quant-ph/0302170}  Y. F. Yu, J. Feng and M. S.
Zhan, quant-ph/0302170.

\end{references}
\end{document}